\documentclass[superscriptaddress,showpacs,amssymb,10pt,reprint,aps,prd,longbibliography,floatfix]{revtex4-2}

\usepackage{graphicx,amssymb,times} 
\usepackage{amsmath,amsfonts}
\usepackage{bm}
\usepackage{hyperref}
\usepackage[usenames]{color}   
\usepackage[dvipsnames]{xcolor}
\usepackage[titletoc]{appendix}
\newtheorem{theorem}{Theorem}

\DeclareMathOperator{\sech}{sech}
\usepackage{graphicx}

\usepackage[normalem]{ulem}

\definecolor{darkblue}{rgb}{0,0,0.5}
\hypersetup{colorlinks=true, citecolor=darkblue, linkcolor=darkblue, urlcolor = darkblue}

\begin{document}

\title{Good tachyons, bad bradyons: \\ role reversal in Einstein-nonlinear-electrodynamics models}

	\author{Marco A. A. de Paula}
%	\email{marco.paula@icen.ufpa.br}
	\affiliation{Programa de P\'os-Gradua\c{c}\~{a}o em F\'{\i}sica, Universidade 
		Federal do Par\'a, 66075-110, Bel\'em, Par\'a, Brazil.}	
		
	\author{Haroldo C. D. Lima Junior}
%	\email{haroldo.lima@ufma.br}
	\affiliation{Departamento de F{\'i}sica, Universidade Federal do Maranh{\~a}o, Campus Universit{\'a}rio do Bacanga, 65080-805, S{\~a}o Lu{\'i}s, Maranh{\~a}o, Brazil.}

	\author{Pedro V. P. Cunha}
%	\email{pvcunha@ua.pt}
	\affiliation{Departamento de Matem\'atica da Universidade de Aveiro and Centre for Research and Development  in Mathematics and Applications (CIDMA), Campus de Santiago, 3810-193, Aveiro, Portugal.}
	
	\author{Carlos A. R. Herdeiro}
%	\email{herdeiro@ua.pt}
	\affiliation{Departamento de Matem\'atica da Universidade de Aveiro and Centre for Research and Development  in Mathematics and Applications (CIDMA), Campus de Santiago, 3810-193, Aveiro, Portugal.}
	
	\author{Lu\'is C. B. Crispino}
%	\email{crispino@ufpa.br}
	\affiliation{Programa de P\'os-Gradua\c{c}\~{a}o em F\'{\i}sica, Universidade Federal do Par\'a, 66075-110, Bel\'em, Par\'a, Brazil.}

\begin{abstract}
In relativistic mechanics, the 4-velocity and the 4-momentum need not be parallel. This allows their norm to have a different sign. This possibility occurs in nonlinear electrodynamics (NED) models minimally coupled to Einstein's theory. Surprisingly, for a large class of NED models with a Maxwell limit, for weak fields, the causal (acausal)  photons, as determined by their 4-velocity, have a spacelike (timelike) 4-momentum, leading to good tachyons and bad bradyons. Departing from weak fields, this possibility is determined solely by the concavity of the NED Lagrangian, which is consistent with the Dominant Energy Condition analysis.
As a corollary, some popular regular black hole solutions sourced by NED, such as the Bardeen and Hayward solutions, are acausal. 
\end{abstract}

\date{\today}

\maketitle

%%%%%%%%%%%%%%%%%%%%%%%%%%%%%%%%%
\textit{\textbf{Introduction.}} 
%%%%%%%%%%%%%%%%%%%%%%%%%%%%%%%%%
In relativistic physics, tachyons are particles with spacelike 4-momentum. On-shell this means they have negative mass squared. In flat spacetime, a tachyonic classical particle is acausal~\cite{Pirani:1970bp}; and a tachyonic classical field has long wavelength (infrared) unstable modes~\cite{Costa:2005ej}. As such, tachyonic particles/modes are a red flag for a physical theory. One famous example is the bosonic string~\cite{Green:1987sp}. 

In non-trivial backgrounds, however, tachyonic particles or modes may become physical. For instance, in Anti-de-Sitter (AdS) tachyons with sufficiently small  (modulus) mass squared are not unstable~\cite{Breitenlohner:1982jf}. Indeed, the size of AdS is an infrared cut-off, avoiding the long wavelength instability. Another example is the phenomenon of spontaneous scalarization~\cite{Damour:1993hw}. The emergence of tachyonic modes reveals an unstable vacuum; their growth drives the configuration to a more stable vacuum, which should exist for sound theories.

\begin{figure}[h!]
		\centering
		\includegraphics[width=0.4\textwidth]{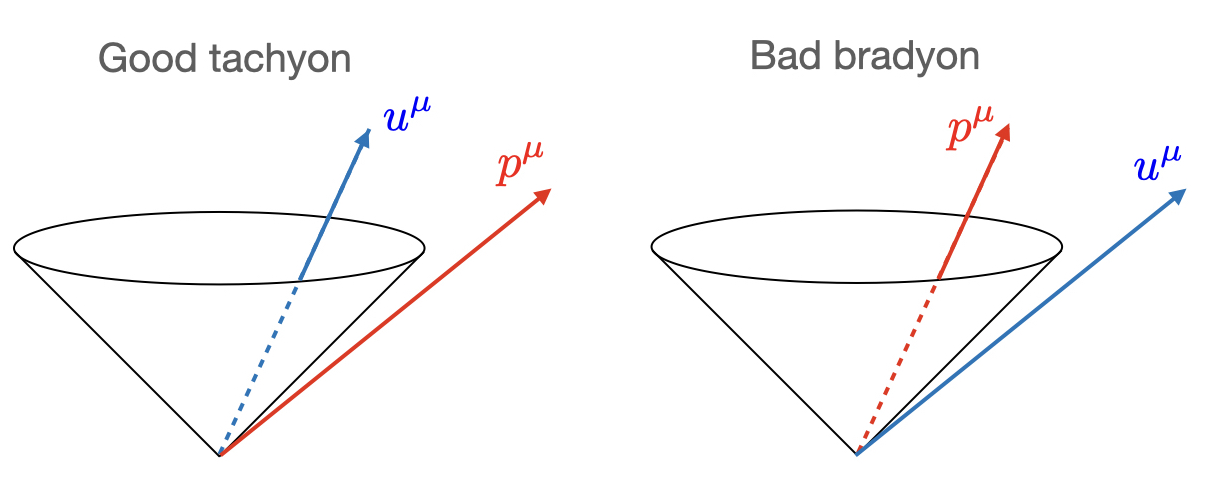}
		\caption{Good tachyons and bad bradyons against the light-cone.}
		\label{fig1}
	\end{figure}

This \textit{Letter} presents a new variant of physical tachyons. In models where modes can have a non-parallel 4-velocity $u^\mu$ and 4-momentum $p^\mu$, tachyonic modes can be causal and bradyonic modes ($i.e.$ with timelike $p^\mu$) can be superluminal and hence acausal. Remarkably, this theoretical possibility is realized in a popular class of models: nonlinear electrodynamics (NED) minimally coupled to Einstein's gravity. In an ironic role reversal, for a class of popular spacetimes in this framework, bradyonic photons are in fact superluminal (or null), whereas tachyonic photons are subluminal (or null) - Fig.~\ref{fig1}.   As a corollary,  some popular black hole (BH) solutions to Einstein-NED theories, such as the Bardeen~\cite{ABG2000} or Hayward~\cite{FW2016} BHs, turn out to be acausal and thus unphysical.

%%%%%%%%%%%%%%%%%%%%%%%%%%%%%%%%%
\textit{\textbf{Unparalleled.}} 
%%%%%%%%%%%%%%%%%%%%%%%%%%%%%%%%%
Let us start with an elementary example where $p^\mu$ and $u^\mu$ are not parallel. Consider a  particle (with mass $m$ and electric charge $q$) in an external electromagnetic field. The canonical conjugate 4-momentum $p^\mu$ relates to the 4-velocity $u^{\mu}$ as~\cite{GPS2001} (we use  natural units, $G = c = \hbar = 1$)
\begin{equation}
\label{classfour}p^\mu=m\,u^\mu+{q}\,A^{\mu}
\ ,\end{equation}
where $A^{\mu}$ is the 4-vector potential, showing $p^\mu$ and $u^\mu$ are not (in general) parallel  and their norms need not match in sign. 

A simple concrete illustration is a relativistic charged particle in a uniform magnetic field along (say) the $z$-axis: the textbook Larmor problem. In a Cartesian chart $(t, x, y, z)$ on Minkowski spacetime, take the 4-vector potential $A^{\mu}=(\phi, \vec{A} )=\left(0, -y\,B+A_0,0 ,0  \right)$,
where $B,A_0$ are constants.  Only the magnetic field is non-trivial, and reads
$\vec{B}=B\hat{z}$.
%\end{align}
Thus $B$ has physical significance, whereas $A_0$ is gauge.
The equations of motion $mdu^\mu/d\tau={q}F^{\mu}_{ \ \ \nu}u^\nu$,
with $F=dA$, are solved by~\cite{Landau_book}
$u^\mu=\gamma(1, u_0\cos\omega\,t, -u_0\sin\omega\,t, 0)$, 
where $\gamma=(1-u_0^2)^{-1/2}$ is the Lorentz factor, $\omega={q\,B}/(\gamma\,m)$ is the relativistic cyclotron frequency, and $u_0$ is the initial condition for the 4-velocity. Neither the electromagnetic field nor $u^\mu$ depend on the gauge constant $A_0$; but 
\begin{align}
\label{classnorm}p^\mu\,p_\mu=\gamma^2\,m^2\left(-1+\omega^2A_0^2+u^2_0\,\sin^2\omega t\right),
\end{align}
depends on $A_0$. 
Choosing $A_0$ sufficiently large, $p^\mu$ can become spacelike, whereas $u^\mu$ is always timelike as long as $u_0<1$. For instance, choosing $A_0=\sqrt{2}/\omega$, then $p^\mu\,p_\mu=\gamma^2m^2\left(1+u_0^2\sin^2\omega\,t \right)>0$. 

This straightforward (yet non-trivial) example in classical electrodynamics highlights that probing the causality of particle motion requires calculating $u^\mu u_\mu$, rather than $p^\mu  p_\mu$. As it turns out, the motion of a photon in NED obeys an equation that shares with this example the non-colinearity between $u^\mu$ and $p^\mu$ - compare Eqs.~\eqref{classfour} and \eqref{dotxmu}. This has been correctly noticed in some previous studies, $e.g.$~\cite{SPL2016}, while others  employed the $p^\mu$ norm as a criterion for causal propagation $e.g.$~\cite{TS2023}. Here, we show that the analysis of the $u^\mu$ norm to assess causality reveals an unexpected twist in the roles of tachyons and bradyons.

%%%%%%%%%%%%%%%%%%%%%%%%%%%%%%%%%
\textit{\textbf{Unparalleled in NED.}}
%%%%%%%%%%%%%%%%%%%%%%%%%%%%%%%%%
Starting from the influential Born-Infeld theory~\cite{BI1934}, NED models have been put forward to generalize Maxwell's theory in the strong field regime. For relativistic invariant models, the most general class is built considering the NED Lagrangian $\mathcal{L}$ to be a general function of the two relativistic invariants constructed from the Maxwell tensor $F_{\mu\nu}$ in four spacetime dimensions. Here, for the simplest possible illustration of the reversal effect, we restrict ourselves to $\mathcal{L}= \mathcal{L}({F})$, where ${F}\equiv F_{\mu\nu}F^{\mu\nu}$ is the Maxwell scalar.

In NED, light rays on a non-trivial background, which could be either an electromagnetic field on Minkowski spacetime or a curved spacetime geometry (or both), follow null geodesics of an \textit{effective} metric~\cite{Boillat:1970gw,P1970,GDP1981,MN2000,Gibbons:2000xe,MP2023}, $\bar{g}_{\mu\nu}$, which in general differs from the spacetime geometry, ${g}_{\mu\nu}$. This phenomenon is analogous to light propagation in the context of nonlinear optics~\cite{NL_OPT}. 
In the eikonal limit, where the associated wavelength is significantly smaller than any characteristic scale of the background field $F_{\mu\nu}$ or geometry $g_{\mu\nu}$, the effective metric is, denoting $\mathcal{L}_{F} \equiv \partial \mathcal{L}/\partial F$ (see Appendix A of Ref.~\cite{SS2019}), 
\begin{equation}
\label{EFF_GEO}\bar{g}^{\mu\nu} \equiv \mathcal{L}_{F} g^{\mu\nu} - 4 \mathcal{L}_{FF} F^{\mu}_{\ \  \sigma}F^{\sigma\nu} \ ,
\end{equation}
on which the photon's 4-momentum $p^\mu$ is null ($\bar{g}^{\mu\nu}{p}_{\mu} \,{p}_{\nu}=0$),  and  geodesic  ($\bar{g}^{\mu\nu}\,{p}_{\mu}\overline{\nabla}_\nu {p}_{\alpha}=0$, where $\overline{\nabla}_\alpha\bar{g}^{\mu\nu}=0$).
 Observe that unless $\mathcal{L}_{FF}=0$, which holds for Maxwell's theory, $\bar{g}^{\mu\nu}$ and $g^{\mu\nu}$ are in general not conformal, and thus photons are not null with respect to the spacetime metric. Notice that this is not simply a consequence of a gauge choice, unlike the apparent tachyonic nature of a charged particle in the initial Larmor problem example.

To relate the photon's 4-velocity and 4-momentum, consider the Hamiltonian governing  photon motion,  
$2\mathcal{H}=\bar{g}^{\mu\nu}{p}_{\mu}{p}_{\nu}$.
Hamilton's equation  $u^\mu\equiv \dot{x}^{\mu} = \partial {\mathcal{H}}/\partial {p}_{\mu}$ yields $u^\mu = {p}_{\nu}\bar{g}^{\mu\nu}$, or, in terms of the spacetime metric,
\begin{align}
\label{dotxmu}u^\mu = \mathcal{L}_F\,{p}^\mu-4\mathcal{L}_{FF}\,h^{\mu\nu}{p}_\nu \ ,
\end{align}
where $h^{\mu\nu} \equiv F^{\mu}_{\ \ \sigma}F^{\sigma \nu}$. 
Equation~\eqref{dotxmu} shares the non-colinear relationship between $p^\mu$ and $u^\mu$ that we have seen for a charged particle in an external electromagnetic field, $cf.$ Eq.~\eqref{classfour}, implying that the norms of $u^{\mu}$ and ${p}^\mu$ may differ in sign. So, let us compute them with respect to the background metric.

 For the 4-momentum we get 
\begin{align}
\label{norm_pa}g^{\alpha\beta}{p}_{\alpha}{p}_{\beta} = \dfrac{4F\mathcal{L}_{FF}}{\mathcal{L}_{F}}F^{\mu}_{\ \  \sigma}F^{\sigma\nu}p_\mu p_\nu \ .
\end{align}
As $g^{\alpha\beta}{p}_{\alpha}{p}_{\beta} \equiv -m_{\text{eff}}^2$ is an invariant, photons in NED can behave as particles with either negative (tachyons),  positive (bradyons) or zero $m_{\text{eff}}^2$, depending on the NED model \textit{and} on the photon motion with respect to the background field. 
On the other hand, for the 4-velocity we get
\begin{align}
\nonumber g_{\mu\nu}\dot{x}^{\mu}\dot{x}^{\nu} = & \left(\mathcal{L}_{F}\right)^{2}g^{\mu\nu}{p}_{\mu}{p}_{\nu}-8\mathcal{L}_{F}\mathcal{L}_{FF}h^{\nu\mu}{p}_{\mu}{p}_{\nu}+\\ 
\label{xuxnug}&\left(4\mathcal{L}_{FF}\right)^{2}h^{\mu}_{\ \ \alpha}h^{\nu \alpha}{p}_{\mu}{p}_{\nu} \ .
\end{align}
The norm of the 4-momentum appears in the first term on the right-hand side of Eq.~\eqref{xuxnug}. The remaining terms can change the relative character (timelike or spacelike with respect to $g_{\mu\nu}$) between $u^\mu$ and $p^\mu$ and determine the true causal properties of the model. We shall now confirm this change is realized in interesting and popular cases, where $u^\mu$ and $p^\mu$ have indeed different characters.

%%%%%%%%%%%%%%%%%%%%%%%%%%%%%%%%%
\textit{\textbf{A class of spacetimes with NED sources.}} 
%%%%%%%%%%%%%%%%%%%%%%%%%%%%%%%%%
Since the late 1990s, BH solutions sourced by different NED models have been discovered~\cite{DPS2021,B2023}. In particular, the observation that regular BHs - singularity-free spacetimes - could be NED sourced~\cite{ABG1998} brought considerable attention to some of these solutions.

The action that describes General Relativity minimally coupled to a NED theory can be written as 
\begin{equation}
\label{S}\mathrm{S} = \dfrac{1}{16\pi}\int d^{4}x \sqrt{-g}\,\left[R-\mathcal{L}(F) \right]  \ ,
\end{equation}
where $R$ is the Ricci tensor. The generalized Maxwell's equations and Einstein equations are
\begin{equation}
\label{EFTC1}\nabla_{\mu}\left(\mathcal{L}_{F}F^{\mu\nu}\right)  = 0 \ \ \text{and} \ \ \nabla_{\mu}\star F^{\mu\nu} = 0 \ ,
\end{equation}  
\begin{equation}
\label{E-NED_F}G_{\mu}^{\ \ \nu} = 8\,\pi\, T_{\mu}^{\ \ \nu} = 2\left[\mathcal{L}_{F}F_{\mu\alpha}F^{\nu\alpha}-\dfrac{1}{4}\delta_{\mu}^{\ \nu}\mathcal{L}(F)\right] \ ,
\end{equation}
where  $\star F^{\mu\nu}$ is the Hodge dual electromagnetic field tensor.

Equations~\eqref{S}-\eqref{E-NED_F} reduce to Einstein-Maxwell theory~\cite{W1984} in the weak field (small $F$) limit if $\mathcal{L}(F) \rightarrow F$ and $\mathcal{L}_{F} \rightarrow 1$. 

The role reversal effect is realized already at the level of simplest (non-rotating) spacetimes. Thus, we consider a static and spherically symmetric spacetime metric
\begin{equation}
\label{LE}ds^{2} = -f(r)dt^{2}+f(r)^{-1}dr^{2}+r^{2}d\Omega^{2} \ ,
\end{equation}
where $d\Omega^{2} = d\theta^{2}+\sin^{2}\theta d\varphi^{2}$ and $f(r)$ is the metric function,  obtained by solving the field equations~\eqref{EFTC1}-\eqref{E-NED_F}. This line element can describe both BHs and horizonless spacetimes.

Here, once more for simplicity, we shall focus on \textit{magnetically} charged \textit{BHs} (or horizonless spacetimes) obtained in the so-called $F$ framework. But our analysis can  be extended straightforwardly to electrically charged BHs obtained in the so-called $P$ framework (a dual variable)~\cite{HGP1987}. For spherically symmetric spacetimes, any magnetic solution obtained in the $F$ framework has an electric counterpart in the $P$ framework and vice versa~\cite{B2001,B2018}. Moreover, the effective geometry of electrically and magnetically charged BHs in the $F$ framework are related by a conformal transformation~\cite{TAM2021}. Consequently, light rays follow the same trajectory in both cases - see  the Appendix of~\cite{NY2022}.

Focusing, therefore, on purely magnetically charged NED sources, the electromagnetic equations can be integrated without knowledge of $f(r).$ The only non-vanishing components of $F_{\mu\nu}$ are $F_{\theta\varphi}=-F_{\varphi\theta}=Q\sin\theta$, where $Q$ is the total magnetic charge~\cite{ABG2000}. Then, the Maxwell scalar reads
\begin{equation}
\label{ms}F = \dfrac{2Q^{2}}{r^{4}} \ .
\end{equation}
Einstein-NED BH solutions of~\eqref{EFTC1}-\eqref{E-NED_F} described by \eqref{LE}-\eqref{ms}, with appropriate $f(r)$ include, $e.g.$, spacetimes we refer to as: Bardeen~\cite{ABG2000}, Born-Infeld (BI)~\cite{AH2022}, Bronnikov~\cite{B2001}, Cadoni et al. (or CLW, for short)~\cite{LW2023,MC2023}, Euler-Heisenberg (EH)~\cite{RWX2013,AA2020} and Hayward~\cite{FW2016}. For completeness, in the Appendix we collect the corresponding NED Lagrangians $\mathcal{L}(F)$.

%%%%%%%%%%%%%%%%%%%%%%%%%%%%%%%%%
\textit{\textbf{Photon motion on magnetic NED sourced spacetimes.}} 
%%%%%%%%%%%%%%%%%%%%%%%%%%%%%%%%%
If the NED source is characterized by a purely magnetic field, with the spacetime metric given by~\eqref{LE}, the corresponding effective geometry line element, $d\bar{s}^{2} \equiv \bar{g}_{\mu\nu}dx^{\mu}dx^{\nu}$, is~\cite{B2001}
\begin{align}
\label{LE_EG2}d\bar{s}^{2} = \dfrac{1}{\mathcal{L}_{F}}\left[ -f(r)dt^{2}+\dfrac{dr^{2}}{f(r)}\right]+\dfrac{r^{2}}{\Phi}d\Omega^{2} \ ,
\end{align}
\begin{equation}
\Phi \equiv \mathcal{L}_{F}+2F\mathcal{L}_{FF}\stackrel{\eqref{ms}}{=}\mathcal{L}_{F}+\dfrac{4Q^{2}}{r^{4}}\mathcal{L}_{FF} \ .
\end{equation}
In models where the Maxwell weak field limit holds, 
\begin{equation}
\mathcal{L}_{F} \rightarrow 1 \quad  {\rm and} \quad \Phi \rightarrow 1 \quad {\rm as} \quad F \ll 1 \ ,
\label{wfl}
\end{equation}
the effective geometry~\eqref{LE_EG2} reduces to the spacetime metric~\eqref{LE} when the limit applies. 

With respect to the effective geometry, the photon's 4-momentum is null. With respect to the spacetime metric, the 4-velocity norm, from Eq.~\eqref{xuxnug}, is
\begin{equation}
\label{velnorm} g_{\mu\nu}\dot{x}^{\mu}\dot{x}^{\nu} = \dfrac{4j^{2}Q^2 \mathcal{L}_{FF}}{r^{6}} \Phi\ ,
\end{equation}
where $j$ is the photon's conserved angular momentum. On the other hand, the 4-momentum norm is
\begin{align}
\label{norm_p}g_{\alpha\beta}{p}^{\alpha}{p}^{\beta} = \dfrac{4j^2Q^2 \mathcal{L}_{FF}}{r^{6}} \left(\dfrac{-1}{\mathcal{L}_{F}}\right) \ .
\end{align}

For a generic NED model, the sign of both  $g_{\mu\nu}\dot{x}^{\mu}\dot{x}^{\nu}$ and $g_{\alpha\beta}{p}^{\alpha}{p}^{\beta}$ depend on  $\mathcal{L}_{F}$ and $\mathcal{L}_{FF}$. For:
\begin{description}
    \item[i)] $g_{\mu\nu}\dot{x}^{\mu}\dot{x}^{\nu} > 0$ ($g_{\mu\nu}\dot{x}^{\mu}\dot{x}^{\nu} < 0$), photons are superluminal (subluminal) with respect to the background light-cone, and hence acausal (causal);
    \item[ii)] $g_{\mu\nu}p^{\mu}p^{\nu} > 0$ ($g_{\mu\nu}p^{\mu}p^{\nu} < 0$), photons are tachyonic (bradyonic) with respect to the background light-cone. 
\end{description} 
Observe that for Maxwell's electrodynamics ($\mathcal{L}_{FF}=0$) or radially moving photons ($j = 0$) in a generic NED model, both the 4-velocity and the 4-momentum are null with respect to the spacetime metric.

Equations~\eqref{velnorm} and~\eqref{norm_p} imply (assuming all quantities involved are non-vanishing)
\begin{equation}
    \frac{g_{\mu\nu}\dot{x}^{\mu}\dot{x}^{\nu}}{g_{\alpha\beta}{p}^{\alpha}{p}^{\beta}}= - \Phi \mathcal{L}_{F}\stackrel{F\ll 1}{\simeq}-1\ ,
\end{equation}
where  the last approximation assumed the existence of a Maxwell weak field limit~\eqref{wfl}. Strikingly the ratio is negative, $i.e.$ velocity and momentum deviate from the background light cone with the \textit{opposite character}, when deviating from the Maxwell limit. We can expect such a behaviour to generically hold beyond the weak field limit, since the ratio would only change sign if either $\Phi$ or $\mathcal{L}_{F}$ changes its sign. However, by continuity, the latter would lead to a singular behaviour in the effective metric~\eqref{LE_EG2}, rendering the photon motion in the NED model ill-defined. We therefore establish the general result, valid as well for electrically charged NED:
\begin{theorem}
\label{theorem0} \textit{Let $\mathcal{L}(F)$ be a NED model admitting a Maxwell weak field limit,  minimally coupled to Einstein's gravity. For static, spherically symmetric, NED sourced spacetime (either electrically and/or magnetically charged), the first order deviation from the Maxwell limit makes photon motion either tachyonic and subluminal, or bradyonic and superluminal, with respect to the spacetime metric. If the NED model is continuous and well-defined everywhere, then the character of the photon motion in the weak field will also hold everywhere else.}
\end{theorem}

%%%%%%%%%%%%%%%%%%%%%%%%%%%%%%%%%
\textit{\textbf{Causality and stability.}} 
%%%%%%%%%%%%%%%%%%%%%%%%%%%%%%%%%
To establish a result without assuming a Maxwell weak field limit, some conditions on $\Phi$ and $\mathcal{L}_F$ can be imposed by invoking the stability of NED-based magnetically charged BHs. We consider the sufficient conditions to have linearly stable BH solutions  outside the event horizon under electromagnetic and gravitational perturbations within our setup. According to~\cite{MS2003,NYS2020}, the linear stability conditions can be written as
\begin{subequations}
\begin{align}
\label{ls1}\mathcal{L}(F) & > 0\ , \\ 
\label{ls2}\mathcal{L}_{F} & > 0 \ , \\
\label{ls3}\Phi & > 0 \ , \\
\label{ls4}3\mathcal{L}_{F} & \geq \Phi f(r) \ .
\end{align}
\end{subequations}

On the one hand, condition~\eqref{ls3} and~\eqref{velnorm} imply the causality of photons is determined solely by the sign of $\mathcal{L}_{FF}$. On the other hand, conditions~\eqref{ls2} and~\eqref{norm_p} imply the tachyon/bradyon character of photons is also determined solely by $\mathcal{L}_{FF}$. Then, for the same sign of $\mathcal{L}_{FF}$, we again establish the two norms have \textit{opposite} signs, this time for any strength of $F$ and without invoking the necessity of a Maxwell limit for the NED model.
This establishes the following:
\begin{theorem}
\label{theorem1} \textit{Let $\mathcal{L}(F)$ be a NED model minimally coupled to Einstein's gravity. A static, spherically symmetric, magnetically charged, linearly stable BH under electromagnetic and gravitational perturbations is causal (acausal) if the NED Lagrangian $\mathcal{L}(F)$ is a concave (convex) function of the Maxwell scalar $F$, i.e., $\mathcal{L}_{FF}<0$ ($\mathcal{L}_{FF}>0$). In the causal (acausal) case photons are tachyonic (bradyonic) or null.}
\end{theorem}

%%%%%%%%%%%%%%%%%%%%%%%%%%%%%%%%%
\textit{\textbf{Illustrations.}} 
%%%%%%%%%%%%%%%%%%%%%%%%%%%%%%%%%
We can confirm these results in concrete cases. Table~\ref{Tab1} presents the sign of $\mathcal{L}_{FF}$ for the aforementioned well-known NED-sourced BHs in General Relativity. For all cases, the  stability conditions~\eqref{ls1}-\eqref{ls4} hold. Observe that the well known Bardeen and Hayward models exhibit acausal photons, despite their everywhere bradyonic (or null) character. These are \textit{bad} bradyons. Indeed, $\mathcal{L}_{FF}$ can take on both positive and negative values in different regions outside the event horizon, depending on the BH's charge-to-mass ratio. However, it will take positive values somewhere.  This questions their (astro)physical viability and the relevance of observability analyses, as the one by the EHT in~\cite{EHT}. 
\begin{table}[hbtp!]
\centering
\begin{tabular}{||c||c|c|c|c|c|c||} 
 \hline
 Geometry & Bardeen & BI & Bronnikov & CLW & EH & Hayward \\ [0.5ex] 
\hline
 $\mathcal{L}_{FF}$ & {\rm Mixed} & - & - & - & - & {\rm Mixed} \\ \hline
Causal & No & Yes & Yes & Yes & Yes & No \\ \hline
DEC & No & Yes & Yes & Yes & Yes & No \\ \hline
\end{tabular}
\caption{$\mathcal{L}_{FF}$ sign, its impact on causality, and abidance of the Dominant Energy Condition (DEC) for some well-known NED-sourced BHs.  The minus sign ``-'' indicate that $\mathcal{L}_{FF}$ is negative. }
\label{Tab1}
\end{table} 

NED-sourced BHs with $\mathcal{L}_{FF}<0$ everywhere outside the horizon also exist, as shown in Table~\ref{Tab1}. These exhibit a causal propagation of light, with the 4-velocity being timelike for non-radial photon motion. Their 4-momentum, on the other hand, becomes spacelike, causing photons to behave like subluminal tachyonic particles; thus \textit{good} tachyons.

%%%%%%%%%%%%%%%%%%%%%%%%%%%%%%%%%
\textit{\textbf{Causality and the Dominant Energy Condition (DEC).}} 
%%%%%%%%%%%%%%%%%%%%%%%%%%%%%%%%%
The (a)causal propagation just discussed should manifest itself at the level of the DEC - see, $e.g.$~\cite{W1984,HE1973,C2019},
which establishes that
\begin{equation}
\label{DEC_def_1}T^{\mu\nu}\xi_{\mu}\xi_{\nu} \geq 0 \ ,
\end{equation}
for all timelike vectors $\xi^{\mu}$, together with the additional requirement that $-T^{\mu\nu}\xi_{\mu}$ is future-directed time-like or null vector.
The DEC is used as a criterion to rule out superluminal motion since energy currents should be causal. 
To assess it in our setup, we consider that the components of the energy-momentum tensor are given by~\cite{HE1973}
\begin{equation}
\label{EMT_PF}[T_{\mu}^{\ \nu}] = \text{diag}(-\rho,p_{r},p_{\perp},p_{\perp}) \  ,
\end{equation}
where $\rho = T_{0}^{0}$ is the energy density. The radial and angular pressures are given by $p_{i} = T_{i}^{i}$ (no sum), for $i = r, \theta, \phi$. Furthermore, spherical symmetry implies that $p_{\theta} = p_{\phi} \equiv p_{\perp}$.

Considering Eq.~\eqref{EMT_PF} and our setup, the set of sufficient conditions to ensure the validity of the DEC for null and timelike trajectories (concerning the spacetime metric) is~\cite{NYS2020}
\begin{subequations}
\begin{align}
\label{PECa_L}\mathcal{L}(F) & \geq 0 \ , \\
\label{PECb_L}\mathcal{L}_{F} & 	\geq 0 \ , \\
\label{PECc_L}\mathcal{L}(F)-F\mathcal{L}_{F} & \geq 0 \ .
\end{align}
\end{subequations}
These equations are obtained independently of the stability conditions [$cf.$ Eq.~\eqref{ls1}-\eqref{ls4}].

In asymptotically flat spacetimes, the Maxwell scalar is vanishingly small at large distances from the BH, as established by Eq.~\eqref{ms}. In this region, we can expand $\mathcal{L}(F)$ as
$\mathcal{L}(F)\approx \mathcal{L}_F(0)\,F+\mathcal{L}_{FF}(0)\,{F^2}/{2}+\mathcal{O}\left(F^3\right)$, 
and substituting in Eq.~\eqref{PECc_L}, we find
\begin{equation}
\label{decimplessuperluminal}-\dfrac{1}{2}F^{2}\mathcal{L}_{FF}(0) \geq 0 \ .
\end{equation}
Therefore,  the DEC demands that  $\mathcal{L}_{FF}(0) \leq 0$ in the far field of the spacetimes in our setup. 
Comparing with~\eqref{norm_p}, in a model with a Maxwell limit, this implies a \textit{tachyonic} photon, which would be intriguing if not for the realization in this \textit{Letter} that the tachyon is subluminal. 

More generically, since the DEC forbids acausal propagation, condition~\eqref{decimplessuperluminal} shows that NED models must have a convex Lagrangian ($\mathcal{L}_{FF}>0$) to allow superluminal photons, corroborating the previous results from the geodesic analysis. This is manifest in Table~\ref{Tab1}. As we can see, the DEC is satisfied (violated) for BH models that are causal (acausal) with respect to the behaviour of the 4-velocity (not the 4-momentum).

Combining the observation that the DEC implies $\mathcal{L}_{FF}\leq0$ with Theorem 2  we can now state the following result:
\begin{theorem}
\label{theorem2} \textit{Let $\mathcal{L}(F)$ be a NED model obeying the DEC, minimally coupled to Einstein's gravity. A static, spherically symmetric, magnetically charged, asymptotically flat, linearly stable BH under electromagnetic and gravitational perturbations is causal, and photons are tachyonic or null.}
\end{theorem}

%%%%%%%%%%%%%%%%%%%%%%%%%%%%%%%%%
\textit{\textbf{Discussions and final remarks.}} 
NED has been a fruitful ground for theoretical explorations, a prime example being the Born-Infeld theory~\cite{BI1934} proposed to resolve the Coulomb singularity of the electric field. It has long been realized that in these models photons 4-momentum is null with respect to an effective geometry, rather than the spacetime one, by virtue of photon-photon interactions in a non-linear theory~\cite{Boillat:1970gw,P1970}. In NED models depending on the two possible relativistic invariants  built from the Maxwell tensor in four spacetime dimensions, $F$ and $G\equiv F_{\mu\nu}\star F^{\mu\nu}$, one can even have two effective geometries, one for each photon polarization, originating a birefringence - see $e.g.$~\cite{Gibbons:2000xe}. 

Despite some solid analyses, $e.g$~\cite{SPL2016,RW2024,MS2024,B2022,SU2011,BT2019}, it has not been sufficiently appreciated, however, that the correct criterion for causality of photon motion is the character of the 4-velocity, rather than the 4-momentum - see $e.g.$~\cite{TS2023} for works using the latter. In this \textit{Letter}, we have provided general criteria to verify whether photon motion is causal in NED-based spacetimes, by investigating the norm of the 4-velocity, given by Eq.~\eqref{velnorm}, which in general differs from that of the 4-momentum, given by Eq.~\eqref{norm_p}.  

For the simplest spacetimes (static, spherically symmetric) and the simplest NED models, depending only on the Maxwell scalar and admitting a Maxwell limit - which is a physical condition -, we observe it is generic that the 4-momentum and the 4-velocity assume a different character (spacelike or timelike). Thus, it is generic that, with respect to the background light cone, tachyonic photons are good and bradyonic photons are bad, in terms of causality.

If one does not wish to assume a Maxwell field limit, it is possible to consider instead a large class of magnetically charged BHs which are linearly stable  - again a physical condition, but not necessarily in NED models with a Maxwell limit. Then, causality depends exclusively on the sign of $\mathcal{L}_{FF}$. And again, tachyonic photons are good and bradyonic photons are bad, in this respect. Moreover, the good models obey the DEC albeit they are tachyonic. Thus, our analysis clarifies this intriguing point.

To illustrate our results, we have considered some well-known BH geometries based with NED sources, namely, Bardeen~\cite{ABG2000}, BI~\cite{AH2022}, Bronnikov~\cite{B2001}, CLW~\cite{LW2023,MC2023}, EH~\cite{RWX2013} and Hayward~\cite{FW2016}. Remarkably, for the Bardeen and Hayward models photons can behave as  bad bradyons, indicating that these solutions are likely unphysical. This illustrates that there are stable BH solutions in the NED framework that do not fulfill the DEC, albeit photons are never tachyonic.

We expect our result to be more general. For instance, introducing rotating in the class of analysed spacetimes will not,  by continuity and at least for small rotation, \textit{per se} change the character of bad bradyons and good tachyons here unveiled. An in-depth analysis for rotating spacetimes, electrically sourced geometries, as well as an extension  to $\mathcal{L}(F,G)$ theories is current underway, to savour a potential extra pinch of surprise. 

Finally, we note that the concept of good tachyons would have striking experimental signatures, even beyond NED. High-energy particles with energy $E$ and 3-momentum $\mathsf{p}$ are known to obey the familiar relation $E^2-\mathsf{p}^2c^2\geq 0$. The discovery of a particle that defies this condition - while still travelling at subluminal speeds — would serve as a definitive ``smoking gun'' for the existence of good tachyons.

\newpage

%%%%%%%%%%%%%%%%%%%%%%%%%%%%%%%%%
\textit{\textbf{Acknowledgments.}} We acknowledge Funda\c{c}\~ao Amaz\^onia de Amparo a Estudos e Pesquisas (FAPESPA), Conselho Nacional de Desenvolvimento Cient\'ifico e Tecnol\'ogico (CNPq) and Coordena\c{c}\~ao de Aperfei\c{c}oamento de Pessoal de N\'ivel Superior (CAPES) -- Finance Code 001, from Brazil, for partial financial support. 
CH thanks the Universidade Federal do Par\'a, in Brazil, while LC thanks University of Aveiro, in Portugal, for the kind hospitality during the completion of this work. 
This work is supported by the Center for Research and Development in Mathematics and Applications (CIDMA) through the Portuguese Foundation for Science and Technology (FCT -- Fundaç\~ao para a Ci\^encia e a Tecnologia) through projects: UIDB/04106/2020, PTDC/FIS-AST/3041/2020, 2022.04560.PTDC (\url{https://doi.org/10.54499/UIDB/04106/2020}; \url{https://doi.org/10.54499/UIDP/04106/2020}; and \url{http://doi.org/10.54499/PTDC/FIS-AST/3041/2020}; \ and \url{https://doi.org/10.54499/2022.04560.PTDC}). This work has further been supported by the European Horizon Europe staff exchange (SE) programme HORIZON-MSCA-2021-SE-01 Grant No.\ NewFunFiCO-101086251.
P.C. is supported by the Individual CEEC program \url{http://doi.org/10.54499/2020.01411.CEECIND/CP1589/CT0035} of 2020, funded by FCT. 

%%%%%%%%%%%%%%%%%%%%%%%%%%%%%%%%%

\textit{\textbf{Appendix.}} Throughout this work, we considered some BH solutions obtained in NED models as sample tests of our main results. In this Appendix, we present the corresponding Lagrangian density $\mathcal{L}(F)$ associated with these models, which are

\begin{itemize}
\item Bardeen~\cite{ABG2000}: 
\begin{equation}
\mathcal{L}(F) = \dfrac{6}{sQ^{2}}\left(\dfrac{y}{1+y} \right)^{5/2},
\end{equation}
\item BI\cite{AH2022}: 
\begin{equation}
\mathcal{L}(F) = 4\beta^{2}\left(\sqrt{1+\dfrac{F}{2\beta^{2}}}-1\right),
\end{equation}
\item Bronnikov~\cite{B2001}: 
\begin{equation}
\mathcal{L}(F) = F \sech^{2}\left[s y^{1/2}\right],
\end{equation}
\item CLW~\cite{LW2023,MC2023}: 
\begin{equation}
\mathcal{L}(F) = \dfrac{F}{\left[1+\left(sy^{1/2}/3\right) \right]^{4}},
\end{equation}
\item EH~\cite{RWX2013,AA2020}: 
\begin{equation}
\mathcal{L}(F) = F-\mu F^{2},
\end{equation}
\item Hayward~\cite{FW2016}: 
\begin{equation}
\mathcal{L}(F) = \dfrac{6}{sQ^{2}}\dfrac{y^{3}}{\left(1+y^{3/2}\right)^{2}},
\end{equation}

\end{itemize}
where $y \equiv \sqrt{Q^{2}F/2}$, $s \equiv |Q|/2M$, $\beta$ is the Born-Infeld parameter, and $\mu$ is a positive dimensionless constant. The parameters $Q$ and $M$ are related to the charge and mass, respectively, of the corresponding central objects.

\end{document}